\documentclass[twocolumn,floats,aps]{revtex4}
\usepackage{times,amsmath,graphics,psfrag,latexsym}

\begin{document}
\title{Surface Curvature and Vortex Stability}
\author{P. Voll, N. apRoberts-Warren, and R.J. Zieve}
\address{Physics Department, University of California at Davis}
\begin{abstract}
We examine the stability of a pinned superfluid helium vortex line
by measuring its persistence at elevated temperatures.  Each vortex
terminates at the surface of the container, at either a rounded bump or
a conical indentation.  We find that pinning with the bump termination
is much easier to overcome thermally.  This behavior would not be
expected from considerations of vortex line tension alone.  We take the
observations as evidence of an additional contribution to the pinning
energetics arising from the interaction of the superfluid order parameter
singularity with the curvature of the container's surface.  By favoring
pinning at points of negative Gaussian curvature, the surface interaction
makes the bump a less advantageous pin site.
\end{abstract}
\maketitle

Topological defects play important roles in a variety of systems.  
Cosmic strings, defects created in the early universe, may provide 
an experimentally accessible test of string theories \cite{Kibble}.
Domain walls affect the behavior of magnetic recording heads and other
magnetic sensors \cite{Spaldin}, and flux line motion in superconductors
introduces dissipation.  Topological defects may also govern protein
folding mechanisms \cite{Nelson}.  In some situations, such as
liquid crystal coatings of small particles, defects must always exist
\cite{Kleman}.  

Recent work has explored how topological defects interact with surface
curvature \cite{VitelliT, VitelliN}, a problem relevant to systems
involving coated particles or flexible membranes.  The energetics
governing the variation of the order parameter favor sites with negative
Gaussian curvature.  Thus point defects confined to a two-dimensional curved
surface tend to position themselves at saddle points.  The same arguments
extend to a three-dimensional system with line defects that terminate
along a curved boundary.

Vortices in superfluid helium are precisely such defects.  Each vortex
core must either close on itself, forming a vortex ring, or terminate at a
surface of the helium.  This surface can be a free surface, if one exists,
or a wall of the container holding the superfluid.  In the latter case,
experimental and computational work shows that the end of the vortex
can become pinned in place.

Early measurements of torsional oscillator damping found a complicated frequency
dependence that can be explained by waves along pinned vortices
\cite{Hall}.  Measurements of thermal counterflow \cite{Hegde} and
rotational acceleration \cite{Adams} also gave evidence for pinning
of vortices on rough walls.  A more direct experimental verification
\cite{pinning} tracked the motion of a single vortex along the cell wall,
including occasional pinning events when the motion ceased.  In some
cases the vortex worked its way free, while in others it remained pinned
until the helium left its superfluid phase.

On the computational side, Schwarz uses a single vortex terminating on a
half-infinite plane and exposed to a constant external flow velocity parallel
to the plane \cite{Schwarz85}.  For sufficiently low flow rates, if the plane
has a hemispherical bump near enough to the vortex's path, the vortex will
spiral onto the bump and remain pinned there. The simulations correspond well
to a naive picture of vortex pinning: by terminating atop the hemisphere, the
vortex has less length and correspondingly less kinetic energy.  In an
alternative formulation, the fluid velocity is highest atop the bump, causing a
reduction in pressure that attracts the vortex to this position.

The present experiment tests the pinning of a vortex by a bump.  Since a
roughly hemispherical bump on a flat background has positive
Gaussian curvature everywhere except near the rim where the bump meets
the flat surface, the predicted curvature interaction \cite{VitelliT,
VitelliN} would favor vortex pinning around the edge of the bump, rather
than at its peak.

We use a straight vibrating wire
to trap a single vortex in superfluid ${}^4$He.  As described elsewhere
\cite{4Heprecess}, we detect the vortex through a change in the
beat frequency of the wire's lowest
normal modes.  The measurements are done on a pumped ${}^3$He cryostat,
which we rotate to create vorticity.  All measurements, however, take
place with the cryostat stationary.  If a vortex becomes trapped along
the entire length of the wire, an especially stable configuration
ensues.  Unless we deliberately disturb the vortex in some way, it
usually remains in place until the cryostat warms to near or above the
superfluid transition temperature.  This behavior was the basis for the
original demonstrations of quantized circulation in superfluids ${}^4$He
\cite{Vinen, WZ} and ${}^3$He \cite{3HeQC}.  With mechanical or thermal perturbation, the vortex
can dislodge from the wire, leading to circulation values intermediate
to the expected quantum levels.  A common configuration after the vortex
comes free is for one end of the vortex to leave the wire and progress
through the cell, terminating on the cylindrical wall
\cite{4Heprecess, helicopter}.  Not surprisingly, such a partially
attached vortex has an intermediate effect on the normal modes.  With a
50 mm wire, we can detect the position of the attachment point in this
configuration to better than 10 $\mu$m precision for a vortex detaching
near the middle of the wire.

\begin{figure}[tb]
\begin{center}
\psfrag{Energy change (10-16 J)}{\scalebox{2.5}{Energy change ($10^{-16}$ J)}}
\psfrag{Distance from wire (mm)}{\scalebox{2.5}{Distance from wire (mm)}}
\psfrag{left}{\includegraphics{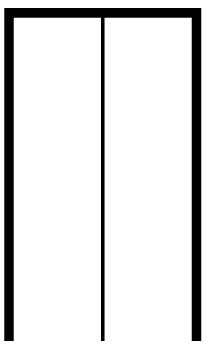}}
\psfrag{mid}{\includegraphics{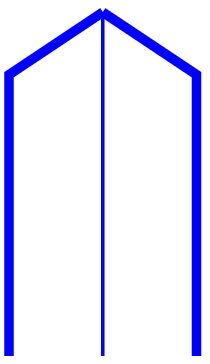}}
\psfrag{right}{\includegraphics{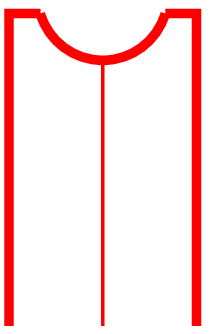}}
\scalebox{.42}{\includegraphics{energies.eps}}
\caption{\small Additional energy as vortex end moves away from wire, if the
end of the cell is flat (solid), conical (dotted), or a rounded bump (dashed).
Insets show these different geometries for the cell end.}
\label{f:energies}
\end{center}
\end{figure}

Our measurements focus on the thermal stability of a trapped vortex line. We
assume that as a vortex comes free from the wire, it moves along the end of the
cell until it reaches the cylindrical cell wall.  If the cell endcap is flat, as
in the leftmost inset of Figure \ref{f:energies}, then the vortex length
increases linearly as it leaves the wire.  Furthermore, since the wire is much
larger than the free vortex core, the energy per length of the vortex is larger
when it is not trapped on the wire.  If the vortex end is a fixed distance from
the wire, minimizing the total vortex energy shows that the detached portion
makes an angle $\theta$ with the wire, where $\cos \theta=\frac{E_W}{E_F}$. Here
$E_F$ and $E_W$ are the energies per length of a free vortex and of a vortex
trapped on the wire, respectively.  Using a free vortex core radius of 1.3 \AA, a
wire radius of 8 $\mu$m, and a cell radius of 1.5 mm gives $\theta=71^\circ$. 
With this value, we can then calculate the change in energy as the vortex end
moves away from the wire, shown as the solid black curve in Figure
\ref{f:energies}.   Once this energy barrier is overcome and the vortex has
reached the cylindrical wall, its motion down the wall decreases the length of
the trapped vortex (and the corresponding energy) without further change to the
average free vortex length.  

Changing the endcap geometry affects the energy considerations.  If the
endcap is drilled out, it resembles the middle inset of Figure
\ref{f:energies}.  The half-angle is $59^\circ$, the angle of the tip of a
standard twist drill.  This geometry reduces the energy price for the vortex to
move away from the wire, since the vortex length increases less than for
a flat endcap.  The dotted blue curve in Figure \ref{f:energies}
shows this new energy.

A third geometry for the cell end, sketched in the rightmost inset of Figure
\ref{f:energies}, is a roughly
hemispherical bump.  Here the geometry enhances the energy as the vortex
leaves the wire.  The dashed red curve in Figure \ref{f:energies} plots
the energy for a hemispherical bump with radius 1 mm.

In a conventional picture of vortex pinning, considering only the line energy
of the vortex, one would expect the energy barrier to the vortex leaving the
wire to be largest when the vortex terminates on a bump, and smallest when it
terminates on a conical surface.  However, the predicted geometric
contribution to vortex energy from surface curvature alters the situation for
cells that end with a bump.  If the geometric term is strong enough, then the
pinned vortex will not follow the wire until it reaches the bump.  Rather, the
vortex will leave the wire and terminate in the negative Gaussian curvature
region around the edge of the bump. From there, the vortex need only traverse
the remaining flat portion of the endcap to reach the cylindrical wall.  Since
the distance involved is shorter than if the wire began in the center of the
endcap, the additional energy required to reach the cylindrical wall is
{\em lower} than for either of the other two end geometries.  This corresponds
in Figure \ref{f:energies} to considering the energy change for the dashed
curve between 1 mm and 1.5 mm, rather than from 0 to 1.5 mm.

The bumps are formed from Stycast 1266.  We begin with a stycast surface
cut flat with an end mill.  To get an aspect ratio near 1 for the bump,
we add a droplet of Stycast 1266 that is already partly set and is
fairly viscous.  Once the bump has dried, we measure its dimensions and
drill a small hole through it for our wire.  After the wire is in place,
we add additional stycast to cover the hole that it emerges from.

The measurements described here come from three cells.  Cell 1, of radius
1.5 mm, has one conical and one flat end.  Cells 2 and 3, of radius 3.5 mm,
each have a bump at one end with the other end flat.  The larger radius in
cells 2 and 3, which was used to provide space for the bump, should
also increase the energy barrier for a vortex in these cells to depin.  In cell
2 the bump has height 1.9 mm and radius 1.9 mm.  Its cross-section along the
cell end is quite regular as well, making it close to hemispherical.  The bump
in cell 3 has height 1.2 mm and radius 1.8 mm, with an irregular
cross-section.  For both cells 2 and 3 the Stycast bumps are not perfectly
centered.  Their closest approach to the cell wall is between 0.5 and 1.0 mm. 
The wire of cell 1 is centered much better, entering the cell through a hole
0.44 mm in diameter at the tip of the conical surface.  Thus part of
each bump is closer to the cell wall than is the wire of cell 1.  

\begin{figure}[b]
\begin{center}
\psfrag{<k> per h/m4}{\scalebox{2.5}{$\langle\kappa\rangle$ ($h/m_4$)}}
\scalebox{.4}{\includegraphics{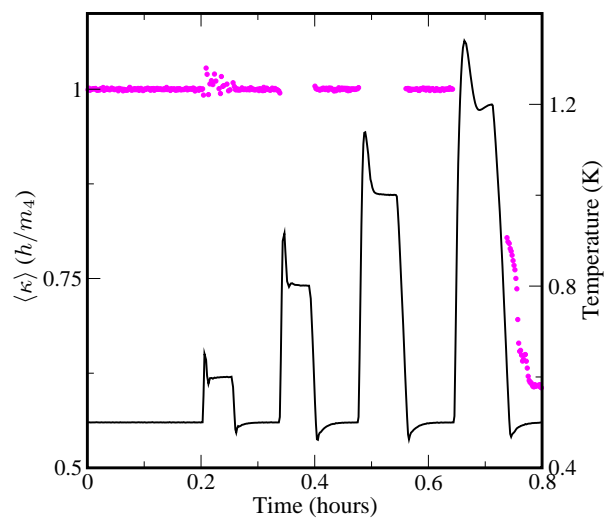}}
\caption{\small Typical temperature sequence (black, right axis) and 
circulation behavior (magenta, left axis) for cell with bump at one end.}
\label{f:heatseq}
\end{center}
\end{figure}

To test stability, we first trap a vortex so that it appears to cover the
entire wire.  We then heat the superfluid to test at what temperature the
vortex comes free.  Since the damping on the wire is too high for reliable
measurements above 600 mK, we cool down after a few minutes to check whether an
end of the vortex has dislodged.  If not, we raise the temperature again,
usually to a slightly higher value, and repeat the cycling until the vortex
does leave the wire.  In some cases the vortex remains trapped until the
cryostat exhausts its ${}^3$He supply and warms above $T_\lambda$. 
Figure \ref{f:heatseq} shows a typical
heating sequence, along with the signal we
observe from the vortex.  The temperature is raised successively to 600 mK, 800
\begin{figure}[!b]
\begin{center}
\scalebox{.7}{\includegraphics{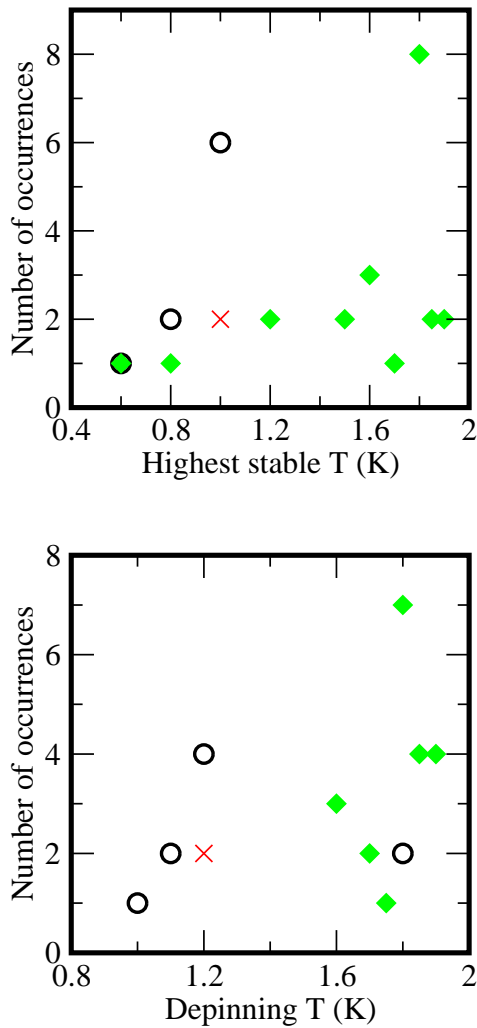}}
\caption{\small Highest annealing temperatures that fails to depin a vortex 
trapped at $\langle\kappa\rangle=1$ (upper) and temperatures that do depin such
a vortex
(lower).  In some cases the cryostat warmed up before the vortex depinned; this
explains some low temperatures in the upper graph.  Similarly, in some cases
the first anneal was at a high temperature and the vortex immediately depinned;
this accounts for some high temperatures in the lower graph.  Data for three
wires are shown; cells 2 (black circles) and 3 (red x's) terminated in bumps
while cell 1 (green filled diamonds) did not.}
\label{f:depin}
\end{center}
\end{figure}
mK, 1 K, and 1.2 K. (The temperature briefly overshoots its target each time,
since the cryostat's temperature control is optimized for rapid settling at a
new temperature.  The overshoots are repeatable from one thermal cycle to the
next.) On cooling back to 500 mK, the vortex clearly remains in place after the
first three anneals, but not after the last. We often find vortex precession
signatures once the circulation level falls below $\langle\kappa\rangle=1$,
which means that the vortex is dislodging from the wire at one end.  The extra
noise in the circulation data at 600 mK comes from the increased damping of the
vibrating wire.  At higher temperatures, the damping is so large that we cannot
extract the circulation at all, so the circulation curve has gaps during the
high-temperature regions.

We repeated this procedure with many trapped vortices on the three wires.  The
results are summarized in the histograms of Figure \ref{f:depin}. The filled
diamonds represent cell 1, the open circles cell 2, and the x's cell 3. The
upper graph shows the highest temperature reached without dislodging the
vortex, while the lower graph shows the temperatures at which vortices did come
free.  The two graphs do not represent exactly the same vortices. Some vortices
on the upper graph never dislodged before the cryostat warmed up; in this case
there is no corresponding point on the lower graph, and the temperature in the
upper graph may be artificially low.  Conversely, some vortices on the lower
graph dislodged on their first annealing.  That anneal may have been above
the minimum
temperature needed to dislodge the vortex, so some points on the
lower graph may be artificially high. The key observation is that in the
absence of the bump no vortex ever depinned below 1.6 K (lower graph), while
with a bump no vortex ever remained pinned above 1 K (upper graph).  This
dramatic difference clearly indicates that vortices along the wire are 
{\em less} stable in the presence of a bump.

In one respect, the histograms do not adequately display the contrast
between the cells.  Each trapped vortex figures only once in each graph
of Figure \ref{f:depin}.  In cell 1, a vortex is often
extremely stable even at the highest annealing temperatures, lasting through 
not just one thermal cycle to 1.8 K but twenty or more.  Thus the difference
in the number of thermal cycles survived by vortices is far greater than
suggested by Figure \ref{f:depin}.

\begin{table}[b]
\caption{Possible relationships among energy barriers to dislodging a vortex.
The first two columns compare the energy barriers at the two ends of each cell. 
The third column incorporates the experimental result that the barrier is 
highest in cell 1.  Terms in parentheses in the third column simply repeat
an inequality from one of the other columns.}
\label{t:endcompare}
\begin{tabular}{l|l||l}
Cell 1 & Cells 2 and 3\hspace*{.3in} & Measurement implication \\
\hline
\hline
cone$<$flat$_1$\hspace*{.3in} & bump$<$flat$_2$ & bump$<$cone\\
\hline
flat$_1$$<$cone & bump$<$flat$_2$ & bump$<$flat$_1$($<$cone)\\
\hline
cone$<$flat$_1$ & flat$_2$$<$bump & flat$_2$$<$cone($<$flat$_1$)\\
\hline
flat$_1$$<$cone & flat$_2$$<$bump & flat$_2$$<$flat$_1$
\end{tabular}
\end{table}

In the present experiment, we cannot directly detect which end of the vortex
dislodges. We expect each cell to have a less stable end at which the vortex
generally works free.  Table \ref{t:endcompare} shows the different
possibilities for where the vortex depins in each cell. Our measurement shows
that the vortex depins more easily in cells 2 and 3 than in cell 1; the
implication for the relative energy barriers is given in the third column of
Table \ref{t:endcompare}.  Note that in the first two cases, the measurements
suggest that a vortex depins {\em more easily} from a bump than from a conical
end.  The second two cases give no information on the relative stability of
the bump and cone; but both lead to the unphysical conclusion that a 
vortex depins from a flat surface more easily in a larger cell, where it
must travel farther.  Thus we conclude that in cells 2 and 3 the vortex
does depin from the bump, and that the energy barrier for this process is
smaller than for depinning from the conical end of cell 1.

Our stability result is the opposite of what would be expected from the simple
energy considerations of Figure \ref{f:energies}.  As suggested above, an interaction between the
vortex and the surface curvature may provide the explanation. Although the
existing calculations consider a two-dimensional system \cite{VitelliT}, we
note that the interaction energy $\frac{\rho_s\kappa^2}{4\pi}V$ is at least
comparable in magnitude to the line energies in our geometry.  Here $\rho_s$
and $\kappa$ are the superfluid density and quantum of circulation,
respectively, and $V$ is a curvature-dependent factor of order 1.

Since our proposed scenario has the vortex pinning to the edge rather than the
top of the bump, we address the possibility of a more direct measurement of
vortex pinning near a bump. If a vortex trapped along the wire terminates at
the edge of the bump, then the vortex must leave the wire shortly before the
wire enters the top of the bump. The short length of wire with no circulation
around it will slightly alter the observed beat frequency. However, the
accuracy of our measurement is much lower than its precision, since an absolute
calibration depends on the mass per length of the wire.  Although we know this
density approximately, it varies by up to 10\% among wires. Each wire is a
single strand of NbTi, approximately 16 $\mu$m in diameter, cut from
multifilamentary superconducting magnet wire. Furthermore, the sensitivity is
much lower near the end of the wire, since the influence of the vortex on the
wire disappears at a vibrational node. In typical measurements one value of
the beat frequency near
where we expect N=1 is far more stable than any others, and we identify this
value with N=1.  We cannot distinguish whether this value corresponds to a
vortex covering the entire wire or only 98\% of it.  We would find a direct
signal only if the vortex sometimes covered the entire wire and sometimes
detached a short distance from the end.  We would then observe more than one
fairly stable circulation level in the vicinity of N=1.  An appropriately
shaped bump, with several metastable positions for a vortex to terminate, may allow
this.

For future measurements, we can determine at which end the vortex comes
free by using a cell with a diameter change in the middle.  After a
vortex works free, its subsequent precession, which depends strongly on
the local cell diameter \cite{helicopter}, will identify which half of
the cell contains the detached portion.  With this information we can
better compare the pinning strength of different endcap geometries.
Another plan for further work is to position a bump on the cylindrical
wall midway along the cell, where the measurement sensitivity is highest.
We have previously seen oscillation signatures for a vortex pinning
on wall roughness \cite{pinning}, and it would be interesting to see
how a vortex behaves on encountering a larger and better-characterized
obstruction.  Quantitative measurements of the pin strength may also
be easier with the improved sensitivity.  Our present results already
demonstrate the need for an additional contribution to the energetics
of a pinned vortex, such as an interaction potential between the vortex
and the surface curvature \cite{VitelliT, VitelliN}.

We thank D. Nelson and V. Vitelli for helpful discussions.  This work
was supported by the National Science Foundation under PHY-0243904 and
by UC Davis.

\end{document}